\documentstyle[a4,12pt]{article}
\begin{document}

\begin{titlepage}

\rule{0em}{5em}

\begin{center}
\LARGE

The continuum limit of the integrable open XYZ spin-1/2 chain

\end{center}

\vspace{2em}

\begin{center}

Hiroshi Tsukahara and Takeo Inami

\vspace{1em}

Department of Physics, Chuo University,\\
Kasuga, Bunkyo-ku, Tokyo 112-8551

\end{center}
\vspace{2em}

\begin{abstract}

 We show that the continuum limit of the integrable XYZ spin-1/2 chain on a 
half-line gives rise to the boundary sine-Gordon theory using the perturbation
method.

\end{abstract}

\begin{center}
\end{center}

\end{titlepage}

\section{Introduction}

 The introduction of boundary interactions to integrable 1+1-dimensional 
quantum field theories have given rise to a series of interesting applications
in various branches of physics ranging from particle physics to 
low-dimensional systems of condensed matter physics with dissipative 
forces or various types of impurities \cite{ludwig:1994,affleck:1995}. 
In view of this, a number of studies have been made to construct 
the integrable extension of both solvable lattice models and 
integrable two-dimensional field theories to the case involving boundaries. 
Clarifying the connection between the former and latter models in the presence 
of boundary interactions is then an interesting question. 
In this letter we focus on the connection of the open XYZ spin-1/2 
chain to the sine-Gordon theory on the semi-infinite line $[0,\infty)$.

 The sine-Gordon theory with a boundary potential depending only on the 
boundary field $\varphi(0)$ is integrable, if the potential is of the form 
\cite{sklyanin:1987,ghoshal-zamolodchikov:1994}
\begin{equation}
V\left(\varphi\left(0\right)\right) = -M\cos\left[\frac{\beta}{2}
\left(\varphi\left(0\right)
 - \phi_0 \right) \right].
\label{eq:boundary_potential}
\end{equation}
Here $\beta$ is the coupling constant appearing in the bulk theory, and $M$ and
$\phi_0$ are arbitrary parameters. The existence of the first few integrals 
of motion in this theory, which we call the boundary sine-Gordon theory, 
can be checked by explicit computations. In the bulk theory, 
the continuum limit of the XYZ spin-1/2 chain is known to be described by the 
sine-Gordon theory \cite{luther:1976}. 
In the lattice theory, the integrable extension of the XYZ model to the case 
with boundary interactions has been made by constructing the reflection matrix 
\cite{inami-konno:1994,guan-tong-zhou:1996}. 
The general form of the Hamiltonian for this integrable XYZ model on a 
semi-infinite chain is given by
\begin{eqnarray}
H &=& H_{XYZ} + H_{b.t.}, \label{eq:hamiltonian} \\[1em]
H_{XYZ} &=& -J \sum_{i=0}^{\infty} \left[ \frac 12 \left( S_i^+S_{i+1}^- 
+ S_i^-S_{i+1}^+ \right) + \Delta
S_i^zS_{i+1}^z + \frac 12 \Gamma \left( S_i^+S_{i+1}^+ + S_i^-S_{i+1}^- 
\right) \right] 
\label{eq:hamiltonian_bulk} \\[1em]
H_{b.t.} &=& A S_0^z + B S_0^- + C S_0^+, \label{eq:hamiltonian_boundary}
\end{eqnarray}
where the external fields $A, B, C$ are arbitrary parameters. In the following 
we assume that $C = B^*$ so that the Hamiltonian possesses the hermiticity.

 In this letter we will show that the continuum limit of the open XYZ spin-1/2 
chain (\ref{eq:hamiltonian}) gives rise to the boundary sine-Gordon theory. 
To derive the continuum limit rigorously, the exact evaluation of the 
low-lying excitation spectrum as well as the ground-state energy of the 
open XYZ spin chain is necessary.
The diagonalization of the Hamiltonian (\ref{eq:hamiltonian}) by means of, 
for example, the generalized Bethe ansatz method is a formidable task 
because of the off-diagonal boundary terms, and the calculation has not 
yet been accomplished. Without the rigorous results, we can still make a 
perturbative analysis in the limit 
in which the strength of $zz$ interaction $\Delta$ and the boundary external 
field $A$, $B$, $C$ are small. 
Note that the X-Y anisotropy $\Gamma$ should be designed to approach zero as 
the lattice spacing $a$ goes to zero, for the continuum limit to be defined.

\section{XYZ model on a line}

 We begin by recapitulating the perturbation method to derive the continuum 
limit of the closed XYZ spin chain \cite{affleck:1990}, 
thereby explaining the tools to be used later in the extension 
to the case of open chain.

 The XYZ spin-1/2 chain can be mapped to the interacting fermions on a lattice 
using the Jordan-Wigner transformation,
\begin{equation}
S_i^- = \psi_i \exp \left(i\pi {\displaystyle \sum_{j=0}^{i-1}} 
\psi_j^{\dagger}\psi_j\right),
\hspace{2em}
S_i^z = \psi_i^{\dagger}\psi_i - \frac 12.
\end{equation}

 We deal with the continuum limit of this model by means of the perturbation to
the XX model, i.e., the limit in which the parameters 
$\Gamma$ and $\Delta$ are small.
The continuum theory of the XX model is that of free massless fermions. 
We express the low-energy excitations in terms of the chiral fermions 
defined on a pair of odd and even sites,
\begin{equation}
\psi_{2l} = (-)^l \left( \psi_{+,\nu} + \psi_{-,\nu} \right),
\hspace{1em}
\psi_{2l+1} = i(-)^l \left( \psi_{+,\nu} - \psi_{-,\nu} \right),
\end{equation}
where $\nu = l + 1/4$. We translate the lattice theory into the continuum 
theory using the prescription: $\psi_{\pm}(x) = 
(2a)^{-1/2}\psi_{\pm,\nu},\, x = 2a\nu$ and ${\displaystyle \int} dx = 2a 
{\displaystyle \sum_{\nu} }$\, , in the limit of zero lattice spacing $a$. 
The resulting Lagrangian (density) of the low-energy effective theory for 
the XYZ model is equivalent to that for the massive Thirring model,
\begin{equation}
{\cal L}^F = 2iv_0:\!\psi_+^{\dagger}\partial_+\psi_+ + \psi_-^{\dagger}
\partial_-\psi_-\!:
+v_1J_+J_- + iv_2 \left( \psi_+\psi_--\psi_-^{\dagger}\psi_+^{\dagger} \right).
\label{eq:thirring}
\end{equation}
Here $J_{\pm}$ are the right and left moving fermion currents, $J_{\pm}
(x^{\pm}) =\; :\!\psi_{\pm}^{\dagger} \psi_{\pm}(x^{\pm})\!:$,\ and 
$x^{\pm} = t \mp x$, $\partial_{\pm} = \left( \partial_t \mp \partial_x 
\right) /2$. 
The parameters appearing in (\ref{eq:thirring}) are related to the coupling 
constants in the spin-chain Hamiltonian (\ref{eq:hamiltonian}) as 
$v_0 = aJ \left( 1 - \pi^{-1} \Delta \right)$,\, $v_1 = 4aJ \Delta$,\, 
$v_2 = J \Gamma$. Setting the value of the spin-wave velocity $v_0$ to 
the unity, we have 
$v_1 = 4\Delta \left( 1 - \pi^{-1}\Delta \right)^{-1}$ and $v_2 = 
a^{-1}\Gamma \left( 1 - \pi^{-1}\Delta \right)^{-1}$. 
To obtain the field theory limit one has to tune the lattice coupling 
constants appropriately as we take the limit $a \rightarrow 0$. 
Thus, we should make the coupling constant $\Gamma$ scale as 
$\Gamma = a\Gamma_r$ where $\Gamma_r$ is a renormalized coupling constant.

 The fermionic field theory can be converted to the bosonic field theory 
through the bosonization,
\begin{eqnarray}
J_{\pm} &=& \mp2\gamma^{-1}\partial_{\pm}\varphi_{\pm},
\label{eq:bosonization-1}\\[1em]
\psi_{\pm} &=& \sqrt{2\mu}\gamma^{-1} e^{\pm i \gamma \varphi_{\pm} }, 
\label{eq:bosonization-2}
\end{eqnarray}
where $\gamma = \sqrt{4\pi}$ and $\mu$ is an infrared cutoff. The chiral boson 
fields $\varphi_+(x^+)$ and $\varphi_-(x^-)$ obey the commutation relations 
at an equal time, 
$[\varphi_+(x^+), \varphi_+(y^+)] = -[\varphi_-(x^-), \varphi_-(y^-)] = 
(i/4) \epsilon(x-y)$ and $[\varphi_+(x^+), \varphi_-(y^-)] = i/4$. 
Introduce the nonchiral boson fields
\begin{equation}
\varphi = \varphi_+ + \varphi_-, \;\;\; 
\tilde{\varphi} = \varphi_+ - \varphi_-.
\end{equation}
They both satisfy the canonical commutation relations and have the commutation 
relation 
$[\tilde{\varphi}(t,x), \varphi(t,y)] = i\theta(x-y)$ between them. 
The Lagrangian 
(\ref{eq:thirring}) then can be represented solely in terms of one of the boson
fields, $\tilde{\varphi}$. After rescaling the fields as
\begin{equation}
\sigma = \varphi/\gamma R, \;\;\; \tilde{\sigma} = \gamma R \tilde{\varphi},
\label{eq:rescale}
\end{equation}
we have the Lagrangian for the sine-Gordon theory
\begin{equation}
{\cal L}^B = \frac{1}{2} \left(\partial_{\mu}\tilde{\sigma} \right)^2 + 
\frac{m^2}{\beta^2} \cos \beta \tilde{\sigma}.
\end{equation}
Here the radius of the boson field $R$ and the mass scale $m$ are related to 
the parameters of the lattice model by 
$\sqrt{4\pi} R = [ 1 + (2\pi)^{-1}v_1 ]^{1/2}$ and $m^2 
= 4\Gamma_r ( 1 + \pi^{-1}\Delta )^{-1}$, respectively. 
$\beta = 1/R$ is the dimensionless coupling constant of 
the sine-Gordon theory. In the present perturbation analysis, 
we have $\beta^2 = 4\pi \left[ \left( 1 - \pi^{-1}\Delta \right)/ 
\left( 1 + \pi^{-1}\Delta \right) \right]$. The exact 
renormalized value for $\beta$ is given by \cite{luther:1976}
\begin{equation}
\beta^2 = 8\pi \left[ 1 - \pi^{-1} \cos^{-1} \left( -\Delta \right) \right].
\label{eq:exact_coupling}
\end{equation}
The above perturbation result reproduces the exact result 
(\ref{eq:exact_coupling}) to first
order in $\Delta$. This is a nontrivial check of the reliability of the 
present perturbation method.

 We note that the interaction term $\left( m^2/ {\beta}^2 \right) \cos \beta 
\tilde{\sigma}$ arises from the X-Y anisotropy as can be seen from the fact 
that $m^2 \propto \Gamma_r$. 
The Lagrangian (\ref{eq:thirring}) contains another term which arises from the 
$zz$ interaction through the Umklapp process\cite{nijs:1981}. 
However, this term is an irrelevant operator in the region of small 
$|\Delta|$ and it can be neglected. 
It becomes marginal at $\Delta = -1$, which is far outside the range of 
the present perturbation analysis.

\section{XYZ model on a half-line}

 Now we proceed to the continuum limit of the open XYZ spin chain 
(\ref{eq:hamiltonian}). We deal with this problem perturbatively starting 
from an unperturbed state in which the bulk is described with a massless boson
field $\tilde{\varphi}$ with the free boundary condition: $\partial_x 
\tilde{\varphi}|_{x=0} = 0$. The $zz$ interaction term and the X-Y anisotropy 
term in the bulk and the boundary external fields are treated as the 
perturbations on this state.
 Since $\partial_t \varphi = -\partial_x \tilde{\varphi}$, the condition means 
$\varphi(t,0) = \varphi_0$, where $\varphi_0$ is a constant with respect to 
the time variable $t$. In terms of the chiral bosons, it reads
\begin{equation}
%\varphi_+(0) = -\varphi_-(0) + \varphi_0.
\varphi_+(t,0) = -\varphi_-(t,0) + \varphi_0.
\label{eq:boundary_condition}
\end{equation}

 We first take into account the boundary perturbation only. 
The continuum representation of the spin operators at the boundary in terms of
the fermions can be written down following the procedure described for 
the bulk case,
\begin{eqnarray}
S^z(0) &=& a^{-1} S_0^z   = J(0) + G(0), \\[0.5em]
S^-(0) &=& a^{-1/2} S_0^- = \psi(0),
\label{eq:boundary_spin}
\end{eqnarray}
where $\psi = \psi_+ + \psi_-$. $J = J_+ +J_-$ is the vector current and 
$G = \psi_+^{\dagger}\psi_- + \psi_-^{\dagger} \psi_+$ is the scalar density. 
The boundary Hamiltonian (\ref{eq:hamiltonian_boundary}) gives the following 
boundary action.
\begin{equation}
S_{boundary} = -\int_{-\infty}^{\infty} dt \left[ A_r\left( J(0) + G(0) 
\right) + B_r \psi(0) + C_r \psi^{\dagger}(0) \right],
\end{equation}
where  $A_r = a^{-1}A$, $B_r = a^{-1/2}B$, $C_r = a^{-1/2}C$ are the 
renormalized value of the boundary external fields. 
It contains the fermion linear terms as well as the fermion bilinear terms. 
Thus, in the bosonized representation, it is naively expected that 
the boundary perturbation contains interactions of the form 
$\cos \gamma \tilde{\varphi}$ as well as $\cos \gamma \tilde{\varphi}/2$ 
\cite{matveev:1995}. We show that only the latter gives rise to a relevant 
boundary interaction.

 In the presence of the boundary, we modify the bosonization rule 
(\ref{eq:bosonization-2}) slightly as \cite{ameduri-konik-leclair:1995}
\begin{equation}
\psi_{\pm} = \sqrt{2\mu}\gamma^{-1} c_{\pm} e^{\pm i \gamma \varphi_{\pm} },
\end{equation}
where $c_{\pm}$ are the zero mode operators defined by
\begin{equation}
c_{\pm} = \frac 14 \exp \left[ \pm \frac{i}{2} \left( \frac{\pi}{2} - 
\gamma \varphi_0 \right) \right].
\end{equation}
Here $\varphi_0$ is the constant operator introduced in the boundary condition 
(\ref{eq:boundary_condition}). 
The relation (\ref{eq:bosonization-1}) remains unchanged. 

 Applying the bosonization rule to the operators at the boundary, we have for 
the vector current
\begin{equation}
J(0) = -2 \gamma^{-1} \partial_t \tilde{\varphi}.
\end{equation}
This term does not contribute to the action after integrating over $t$. 
For the scalar density term, we see that the two terms of $G(0)$ add up to 
zero. Thus, the diagonal part of the boundary interactions, 
which are quadratic in the fermions, gives no contributions 
in the continuum limit. It remains to evaluate the off-diagonal terms. 
we have
\begin{eqnarray}
B_r\psi(0)+C_r\psi^{\dagger}(0) = \sqrt{2\mu} \gamma^{-1} |B_r| \Bigl\{ 
\left[ \left( c_+ e^{i\gamma\varphi_+(0)} 
+ c_- e^{-i\gamma\varphi_-(0)}\right) \right] e^{-ib} \nonumber\\[0.5em]
+ \left[ \left( e^{-i\gamma\varphi_+(0)} c_- 
+ e^{i\gamma\varphi_-(0)} c_+ \right)
\right] e^{ib} \Bigl\}, \rule{2cm}{0cm}
\label{eq:boundary_fermion_linear}
\end{eqnarray}
where we have put $\arg B_r = -b$. 
 We should recall that it is the field $\tilde{\varphi}$ that is relevant 
in the bulk sine-Gordon theory. 
The boundary interaction (\ref{eq:boundary_fermion_linear}) involves both 
operators $e^{\pm i \gamma \varphi(0)/2}$ and 
$e^{\pm i \gamma \tilde{\varphi}(0)/2}$. 
The free boundary condition implies that the boundary value of the boson field
$\varphi$ is a constant in time, i.e., it is not a dynamical variable. 
Hence, the boundary interaction is also given by the field $\tilde{\varphi}$.

 To obtain the expression in terms of the nonchiral bosons, we first note that 
we have
\begin{eqnarray}
\tilde{\varphi}(0) &=& 2 \varphi_+(0) - \varphi_0, \nonumber\\
                   &=& -2 \varphi_-(0) + \varphi_0,
\label{eq:boundary_bosons}
\end{eqnarray}
at the boundary. Taking care of the fact that the zero mode operator 
$\varphi_0$ does not commute with the boson field $\tilde{\varphi}$, 
we can express the linear combination of the exponentials of the 
chiral bosons at the boundary in terms of $\tilde{\varphi}(0)$ 
\cite{ameduri-konik-leclair:1995}
\begin{eqnarray}
\lefteqn{c_+ e^{+i\gamma\varphi_+(0)} e^{-ib} + e^{-i\gamma\varphi_+(0)} 
c_- e^{ib} } \nonumber\\[0.5em]
 &=& c_- e^{-i\gamma\varphi_-(0)} e^{-ib} + e^{+i\gamma\varphi_-(0)} c_+ e^{ib}
 \nonumber\\[0.5em]
 &=& \frac 12 \cos \left( \frac{\gamma}{2} \tilde{\varphi}(0) - b \right) 
%\rule{18em}{0em}.
\label{eq:formula}
\end{eqnarray}
 Now we include the $zz$ interaction term, which yields merely the rescaling 
(\ref{eq:rescale}) of the boson fields in our perturbation scheme as mentioned
earlier. Substituting the relations (\ref{eq:formula}) into 
(\ref{eq:boundary_fermion_linear}) and rescaling the boson fields, we 
finally arrive at the boundary Lagrangian
\begin{equation}
{\cal L}_{b.t.} = -M \cos \frac{\beta}{2} \left( \tilde{\sigma}(0) - \phi_0 
\right),
\label{eq:boundary_lagrangian}
\end{equation}
where
\begin{equation}
M = \sqrt{2\mu} \gamma^{-1} |B_r|, \;\;\; \phi_0 = \frac{2b}{\beta}.
\end{equation}
Thus the integrable boundary potential (\ref{eq:boundary_potential}) for the 
sine-Gordon theory has been derived from the continuum limit of the integrable
boundary interactions for the open XYZ spin-1/2 chain.
We see that the free parameters $M$ and $\phi_0$ appearing in the boundary 
potential (\ref{eq:boundary_potential}) are related to the strength and 
the phase of the off-diagonal part of the boundary external field for the open
XYZ spin chain.

 Finally, we check the scaling dimension of the operator 
(\ref{eq:boundary_lagrangian}).
To this end, we apply the technique using the analytic continuation 
\cite{eggert-affleck:1992}. 
We make use of the boundary condition (\ref{eq:boundary_condition}) 
to extend the right-moving boson $\sigma_+$ to the negative axis, 
$x < 0$, by defining it as the analytic continuation of the left-moving 
boson $\sigma_-$ with
\begin{equation}
\sigma_+(t,x) = -\sigma_-(t,-x) + \gamma \varphi_0 / \beta \;\;\; (x < 0).
\end{equation}
It is noted that we are now considering the situation in which the bulk is 
still described with a massless free boson $\tilde{\sigma}$, i.e., 
the X-Y anisotropy is not included. This is sufficient to calculate the 
scaling dimension of the boundary operator in our perturbative analysis. 
Since the interactions exist only at the boundary, 
the chiral bosons $\sigma_+$ and $\sigma_-$ are decoupled in the bulk and 
the locality of the theory is not violated under this extension 
\cite{affleck-ludwig:1994}. 
Now the right-moving boson $\sigma_+$ is defined on the whole line 
$-\infty < x < \infty$ and the calculation of the scaling dimensions 
of the operators at the boundary follows from the knowledge in the 
bulk theory. The scaling dimension of the operator 
(\ref{eq:boundary_lagrangian}) is evaluated to be $\beta^2/8\pi$. 
If we exploit the exact value of $\beta$, we see that this operator 
is relevant for all values of $\Delta$ on the critical line, 
$-1 \le \Delta \le 1$, except for the antiferromagnetic point $\Delta = -1$ 
at which it becomes marginal.
Thus, the interaction (\ref{eq:boundary_lagrangian}) indeed survives 
in the continuum limit.

 To summarize, we note that our analysis has clarified that the boundary 
potential (\ref{eq:boundary_potential}) is, in fact, the general solution 
for the boundary potential which is compatible with the integrability of 
the sine-Gordon theory with boundaries, though it has less number of free 
parameters than the general solution for the integrable open XYZ spin chain 
\cite{inami-konno:1994}.

\vspace{2em}
\begin{flushleft}
{\bf Acknowledgements}
\end{flushleft}

 The authors are indebted to Satoru Odake for many helpful discussions 
and valuable comments on the manuscript.

\end{document}